Evidence and concerns about a latent, embryonic phase tectonic evolution and the existence of the young subsurface ocean on Mimas

Balázs Bradák* and Motoharu Okumi

Laboratory of Exo-oceans, Faculty of Oceanology, Kobe University, 5-1-1 Fukaeminami-machi, Higashinada-ku, Kobe 658-0022, Japan

*Corresponding author: B. Bradák, bradak.b@port.kobe-u.ac.jp

Abstract

The icy moons of outer Solar system gas and ice giants have been in the spotlight of scientific and common interest because of the possibility of a subsurface ocean hidden under their ice shells, which may harbor extraterrestrial life. The patterns of various lineaments on the surface of icy satellites may indicate active tectonic processes and the interaction between the surface and the subsurface ocean, which allows material transport toward the habitat of putative alien life. In the case of Mimas, one of the moons of Saturn, new models challenge the long-standing conclusion about the satellite being an inactive snowball, suggesting the existence of a young stealth ocean. Unfortunately, no observable evidence has been found yet implying tectonic activity and the theoretical subsurface ocean. Here, we present the first structural geological map of the icy satellite, with the signs of various tectonic features, along with a simple crosscutting chronology of lineaments formation. In accordance with the supposedly young age of the stealth ocean, the observed phenomena are described as putative lineaments, ridges, and troughs. Such simple tectonic features are identified as young compared to complex structures, such as bands appearing on other satellites. The pattern of the linear features seems to overlap with the allocation of various modeled global nonlinear tidal dissipation patterns. In such a way, it may provide the first observed evidence for the existence of the theoretical subsurface stealth ocean. With such an evolving young subsurface ocean and the barely recognizable pattern of simple lineaments, Mimas may represent a new group of icy satellites showing the early, latent, embryonic phase of tectonic activity. Such a step in the geological evolution of icy satellites has never been observed before. In contrast, the overlapping and crosscutting relation between craters and the observed features may raise some concerns about the "recent" formation of such linear features, indicating possibly long-time dormant or already stopped tectonic processes at the very early, embryonic phase of lineament formation billions of years ago. The presented geological investigation brought a new angle and additional evidence about a possible stealth ocean hiding under the crater-covered surface of Mimas, regardless of its geological age and recent state of evolution. Undoubtedly, the results and the raised concerns will trigger more intense investigations on Mimas and other, similar icy satellites in the Solar system.

1. Introduction

Icy satellites, the moons of outer Solar system gas (Jupiter and Saturn) and ice giants (Uranus and Neptune), have been in the spotlight of scientific and common interest because of the possibility of a subsurface ocean hidden under their ice shells, which may harbor extraterrestrial life. The patterns of various lineaments on the surface of icy satellites, like Europa or Enceladus, may indicate active tectonic processes and the interaction between the surface and the subsurface ocean, supporting the transportation of key materials (oxidants) toward the habitat of potential extraterrestrial lifeforms (Howell and Pappalardo, 2020).

Compared to those icy satellites with a possible subsurface ocean (Nimmo and Pappalardo, 2016), Mimas has been considered an inactive moon, with no characteristic surface indication of global tectonism and the mark of any related surface renewal processes. Backing such old age up, the crater counting method showed 4.3 Ga absolute age at the heavily cratered regions and dated the impact and the creation of the Herschel crater back to 4.1 Gyr ago (Schmedemann and Neukum, 2011). Such commonly accepted knowledge about Mimas changed recently when a series of studies discussed the possibility of a subsurface "stealth" ocean below the frozen and ancient-looking surface of the satellite (Tajeddine et al., 2014; Neveu and Rhoden, 2017; Noyelles, 2017; Rhoden et al., 2017; Rhoden and Walker, 2022; Rhoden, 2023; Ćuk and Rhoden, 2024; Lainey et al., 2024). Although the appearance of a subsurface ocean looks plausible, there are still some controversies about the formation of a subsurface ocean and the lack of any characteristic mark of (cryo)tectonic processes on the surface, which indirectly may indicate the appearance of such liquid layer below the ice crust. Based on various models, the lack of such surface features is explained in many ways, including a strong ice shell, which may withstand higher tidal stresses (Rhoden et al., 2017). Alternatively, the theoretical young geological age of the subsurface ocean, suggested by Rhoden (2023), Ćuk and Rhoden (2024), and Lainey et al. (2024), may explain the lack of tectonic features, implying that if the subsurface ocean and global tectonism exist, the tectonic evolution of the shell is "in progress," appearing in the form of a latent, very early "embryonic phase." Such latent, early-stage tectonism would partly explain the reason why even stagnant lid tectonism-related features (generally showing initial tectonic activity in planetary bodies) are still not recognizable on the ice crust of Mimas (Multhaup and Spohn, 2007).

Two pioneering studies provided the first geological maps of Mimas, considering the knowledge about the moon in the 1990s (Stooke, 1989; Croft, 1991). Since those Voyager-image-based maps, no attempt has been made until recently, when the first version of Mimas's Cassini image-based global geological map was presented (Bradák and Okumi, 2024). Although the early geological maps introduced by Stooke (1989) and Croft (1991) already showed the existence and possible tidal-force-related origin of various linear features on the surface, no systematic mapping and analysis have been made of those features since the first publication of the Cassini data-based Mimas image mosaic map in 2005 (PIA07779; https://science.nasa.gov/resource/map-of-mimas-december-2005/). The rising attention to the moon, triggered mainly by the model-based theory about a subsurface ocean, set off further planetary studies on the satellite's surface. Completing the executed simulations with observations and recognizing any mark of active or inactive stress fields on the surface of Mimas feels essential as evidence of the subsurface ocean and the origin and evolution of the satellite itself. The potential marks of early phase tectonism would support the theory about the "ring origin" of the moon instead of the primordial accretion theory (Rhoden, 2023).

Considering the very recent peak in the scientific and public attention about the existence of an evolving, young ocean under the ice shell of Mimas, this study aims to react to the new theory(ies) and, along with additional comments, introduce some key characteristics of the global tectonic activity on Mimas which may provide further information about the existence and evolution of the putative subsurface ocean as well.

*Please note that the goal of the study is to provide some idea (even without presenting detailed modeling and simulations) as a reaction to the recent development in the study of the icy satellite and provoke further conversations, debates, and versatile research, completing the existing theory about Mimas` young, evolving subsurface ocean.*

## 2. Data and Methods

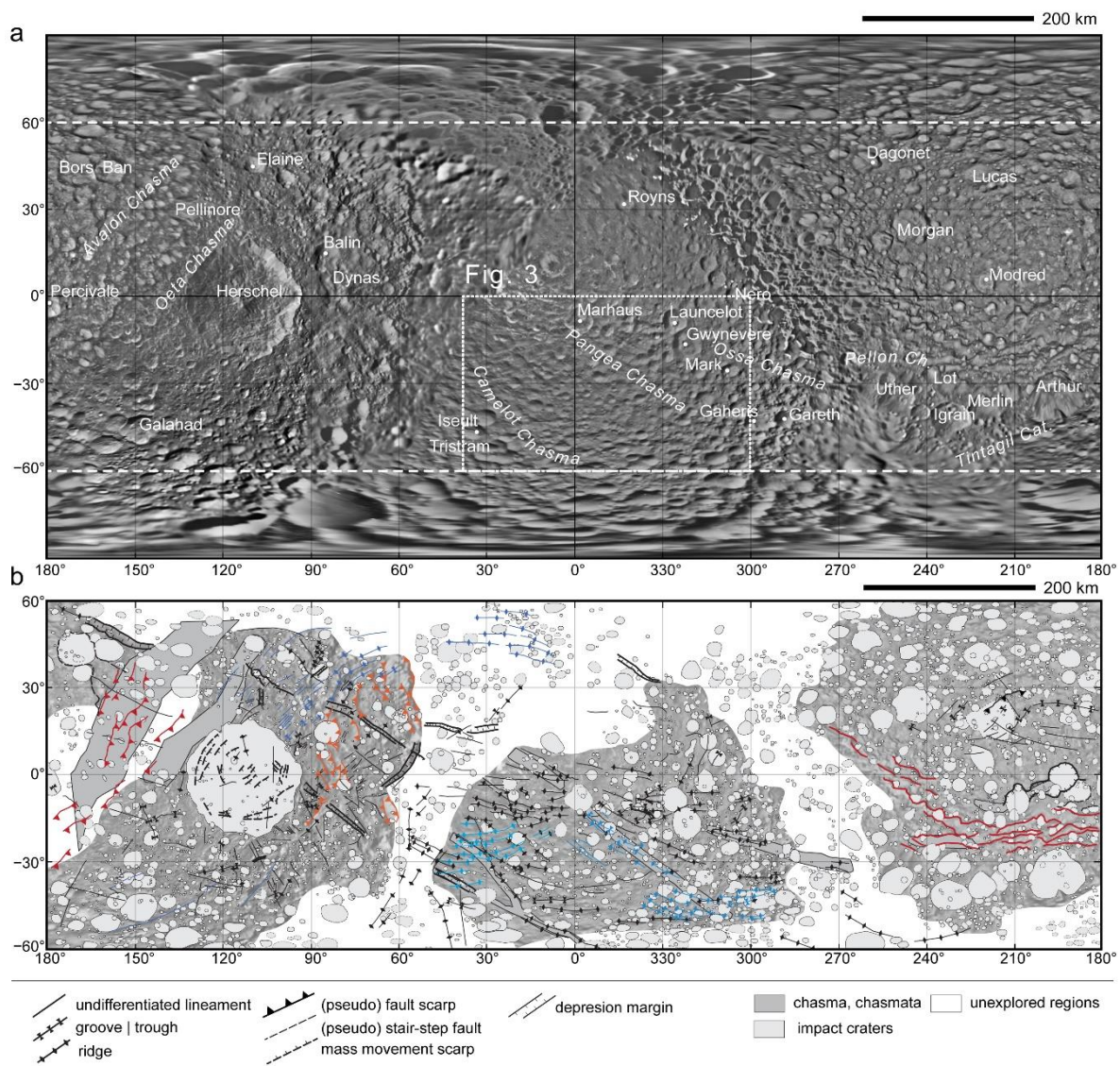

**Figure 1**. The photomosaic was used during the geological mapping (a) and the delivered global tectonic map (b). a) utilizes the Mimas Global Map – June 2017 version, which can be found on the site of NASA Jet Propulsion Laboratory, California Institute of Technology (https://www.jpl.nasa.gov/images/pia17214-mimas-global-map-june-2017), as its base map. The applied nomenclature adheres to the recommendation of the Gazetteer of Planetary Nomenclature (https://planetarynames.wr.usgs.gov/Page/MIMAS/target). Because of the distortion caused by the projection system around the polar regions, the regions beyond latitude +/-60° were excluded from the analysis in Figure 1b, as shown by the dashed white lines. The area inside the dotted rectangle is analyzed in Figure 3. The tectonic features marked by various colors indicate various tectonic processes, such as orange – Herschel impact event-related scarps, red – Ithaca Chasma analog features, and different tones of blue – quasi-parallel linear features, selected for further analysis (please find more information in the text at Section 3, and in Figures 2 and 3).

During the mapping of the putative tectonic features on Mimas, three base maps have been used, namely Mimas Global Map (2017; PIA17214; https://www.jpl.nasa.gov/images/pia17214-mimas-global-map-june-2017), Global 3-Color Map of Mimas (2014; PIA18437; https://science.nasa.gov/resource/color-map-of-mimas-2014/), and a supposedly earlier, version of the former Mimas Global Map (2012; PIA14926; https://photojournal.jpl.nasa.gov/catalog/PIA14926) and appears in JMars (version 5.3.15.2; https://jmars.asu.edu/) named Cassini ISS Cartographic Map of Mimas. Maps and descriptions from the earliest pioneering publications about Mimas geology have also been used to identify various features (Stooke, 1989; Croft, 1991).

The nomenclature used in this study follows the recommendation of the Gazetteer of Planetary Nomenclature (https://planetarynames.wr.usgs.gov/Page/MIMAS/target).

The geological mapping and related GIS research were performed by QGIS 3.22 and JMars 5.3.15.2 software, where all features, including craters and structural elements, were interpreted visually and digitized interactively from the high-resolution and georeferenced images.

The "relative crosscutting age" (RA) was determined based on the crosscutting relationship of various lineaments. RA shows the morphostratigraphical relation between the chosen lineaments following the order of their formation. In this study, the youngest features (with no overlying lineaments) were marked by RA1, and the oldest lineaments, overlaid by many lineaments, were defined up to RA6, indicating their relative crosscutting age (Section 3). More information with a detailed description of the determination of the relative crosscutting age and the potential uncertainties and bias during the analysis can be found in Bradák et al. (2023).

3. Results and Discussion

Around five hundred (501) (quasi-)linear features were identified during the geological mapping of the moon and classified into three main categories such as lineaments, scarps, and mass movement-related features (Bradák and Okumi, 2024). Many features of the latter two categories seem connected to various processes related to the Herschel global-scale impact event (Bruesch and Asphaug, 2004; Moore et al., 2004). The formation of a particular type of landforms and tectonic features may be triggered by the seismic wave of the impact, the antipodal effect, or the appearance of concentrical rings around ground zero (Moore et al., 2004). The pattern of Avalon Chasma and its newly recognized section comprises stair-step faults observed in the neighborhood of Modred impact crater (Fig. 1b) (Stooke, 1989) looks similar to the pattern of Ithaca Chasma, a landform observed on Tethys and possibly related to the Odysseus crater forming impact. If such an analogy is accurate, the formation of the identified fault system can be explained by the whole-body oscillation of the moon triggered by the Herschel event (Moore and Ahern, 1983; Bradák and Okumi, 2024). The development of such tectonic features might be part of forming a ring-graben structure during the impact event, triggered by the collapse of the crater floor, involving materials from the moon's interior and resulting in the collapse of the crater wall, e.g., indicated by the identified mass movements (Schenk, 1989; Bradák and Okumi, 2024).

Regarding the former (lineaments category), earlier studies, referring to the lack of various tectonic features similar to the ones on Europa's surface, suggest that the strong ice crust of the satellite may be capable of withstanding higher tidal stress (Rhoden et al. 2017). The pioneering geological studies of Mimas, conducted by Stooke (1989) and Croft (1991), and the introduced Cassini image-based geological map (Bradák and Okumi, 2024) show different situations. Although the identified features are far less characteristic than the ones on icy satellites with tectonized surfaces (e.g., Enceladus, Europa, Dione, and Ganymede), the basic characteristics and the distribution of those simple, barely identifiable features can help to describe the forming processes and environment. Most of the linear features observed on the surface of Mimas belong to undifferentiated lineaments, ridges, and troughs (grooves), which, considering the theory of Prockter and Patterson (2009), would imply an early stage of lineament formation and supposedly tectonic evolution. The appearance of such latent, "embryonic phase" tectonic activity, along with the heavily cratered, old-looking surface of the satellite (Schmedemann and Neukum, 2011) and the expectedly thick crust at the time of Herschel impact, would fit the theory of a thinning ice shell (<30 km) and the young, evolving subsurface ocean (Denton and Rhoden, 2022), along with the influence of tidal forces appearing in the icy shell (Rhoden et al., 2017, Rhoden and Walker, 2022; Rhoden, 2023). Such a late birth of a subsurface ocean would not only explain the appearance of simple linear features on the crater-covered old surface but may imply a ring origin of Mimas, favored over the primordial accretion in the circum-planetary disk around the gas giant (Rhoden, 2023).

Despite this exciting-sounding idea, there are some additional observations to consider. Instead of appearing as individual ones, many of the simple features appear as a group of quasi-parallel lineament groups (Fig. 1b). Because of the crater coverage and the frequently appearing low-resolution image mosaics, it is challenging to recognize their morphology clearly, but some analogy to lineated and ridged bands (Europa) and groove lanes (Ganymede) feels reasonable to mention here (Howell and Pappalardo, 2018).

In such a way, the existence of those quasi-parallel trough-ridge structures may pinpoint the thinning of the crust (Howell and Pappalardo, 2018) and periodic dilatation-compression cycles ("tidal squeezing"; Dameron and Burr, 2018). Their appearance may point toward developing more complex, band-like structures (Prockter and Patterson, 2009; Howell and Pappalardo, 2018).

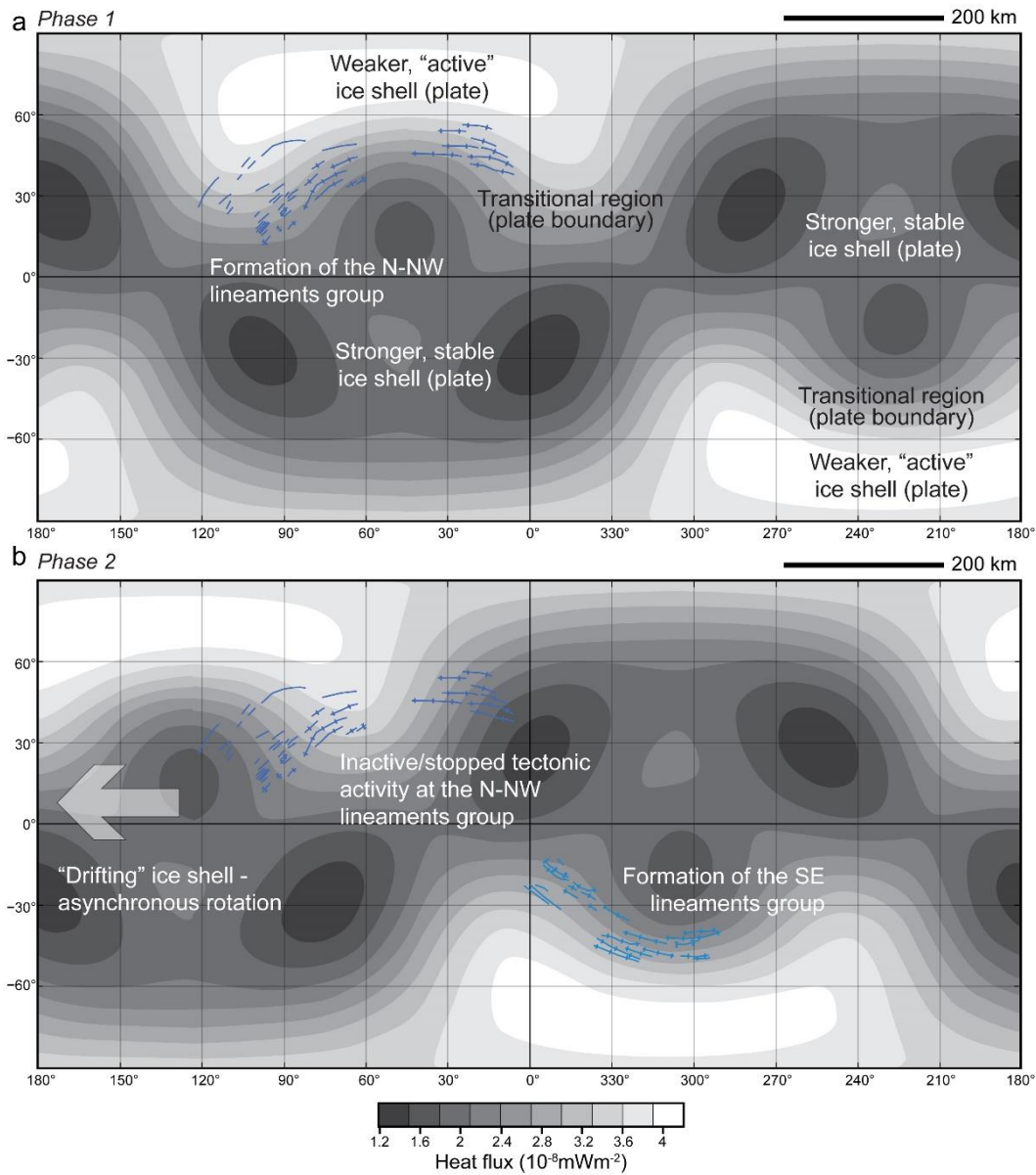

**Figure 2**. The comparison of the allocation and pattern of some selected quasi-parallel lineament groups and the simulated patterns of tidal heating related to the total dynamic forcing of eccentricity and obliquity. The heat flux map is modified after the study by Hay and Matsuyama (2019). Figures 2a and b represent two phases of the tectonic evolution of the ice shell, related to the asynchronous rotation of the ice shell to the tidal torques. Please note that the purpose of the drifting orientation is only to illustrate and explain the theory introduced in Section 3 without considering the rotation and orbital components of the icy satellite.

Comparing the allocation and pattern of some of the lineament groups (Fig. 2a and b) with the simulated patterns of tidal heating related to the total dynamic forcing of eccentricity and obliquity (Quillen et al., 2016; Matsuyama et al., 2018; Hay and Matsuyama, 2019), may provide some additional information. Two remarks feel necessary to mention based on the observation of Figure 2.
- The allocation of the studied feature seems to overlap the zone between the thermally active and least active (inactive) zones, and
- The pattern of heat flux and tidal dissipation on the surface seems to "move," appearing in different regions over time (Fig. 2a and b), considering the relation between the quasi-parallel lineaments and the heat flux pattern.

One interpretation of the former remark is that tectonically active areas are found between two distinct regions, possibly called "global ice plates." One global ice plate is characterized by a more robust and stable crust (low heat flux), and the other is defined by weaker, less stable, more mobile crusts (higher heat flux).

The latter remark suggests a possible indication of an asynchronous rotation of the ice shell to the tidal torques. Such asynchronous rotation may be recognized by comparing the theoretical pattern in Figure 2a and Figure 2b. In Figure 2a, the northwestern lineament group seems to fit the heat flux/tidal dissipation pattern, but the other lineament groups do not. In contrast, the pattern fits the parallel lineament group located southeast after drifting the heat flux/tidal dissipation map (i.e., "simulating" the asynchronous rotation between two components in the satellite's interior). Such observation implies that potentially overlapping generations of lineaments may be found on Mimas surface.

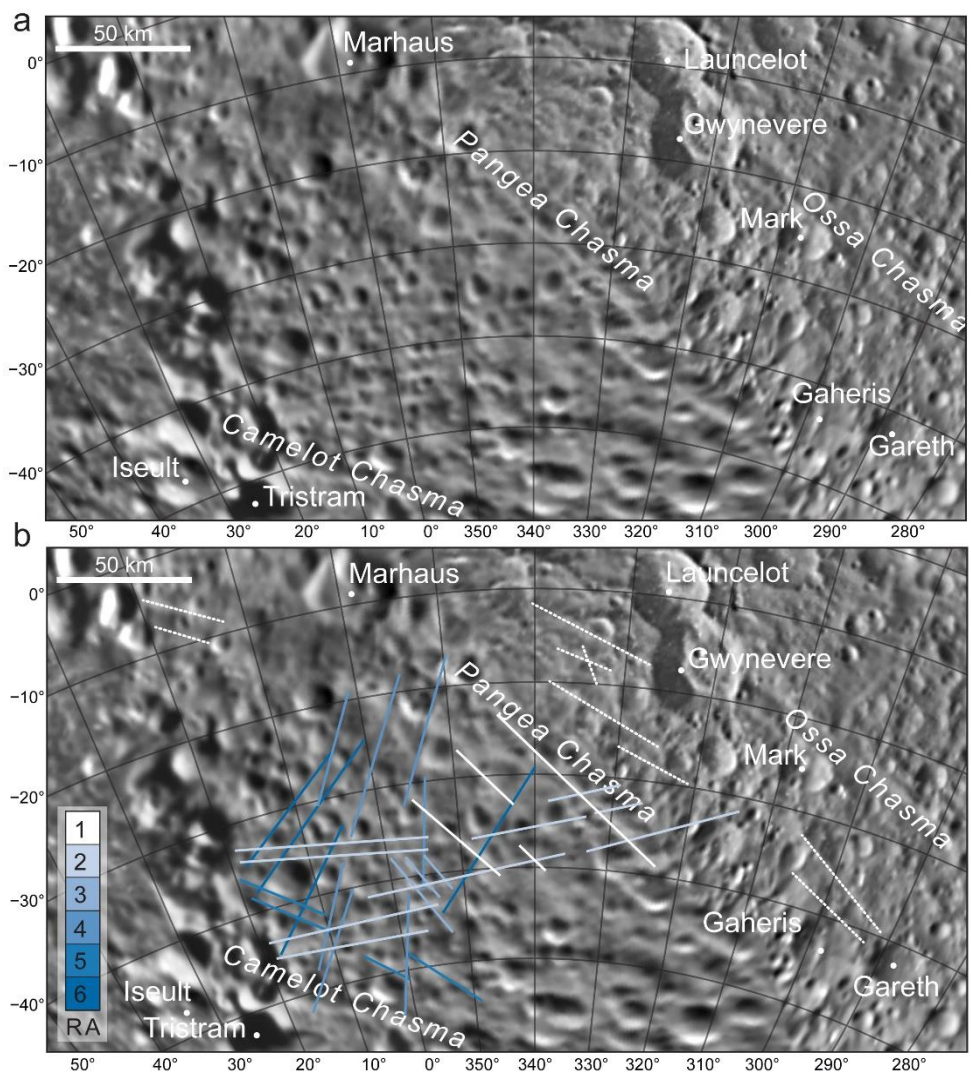

**Figure 3**. Revealing the crosscutting relationship between quasi-parallel lineament groups. Figure 3a shows the original image mosaic with weak, barely visible crosscutting lineaments in the center. Figure 3b shows their crosscutting relation and relative morphostratigraphic age (RA1 to 6). The strike of the identified lineament groups are RA1$_{AVG}$: 132°, RA2$_{AVG}$: 85.8°, RA3$_{AVG}$: 147.9°, RA4$_{AVG}$: 21.8°, RA5$_{AVG}$: 127.5°, and RA1$_{AVG}$: 42.2°. The map is in a recentered projection compared to the original image mosaics and maps in Figure 1. RA colors mark the relative crosscutting age from the youngest, indicated by white (RA1), to the oldest, marked by dark blue

(RA6). The dotted white lineaments show the allocation of additional lineaments on the map, with unknown RA, considering the age of the dated lineaments.

One potential area with overlapping parallel lineaments was found between Chamelot and Pangea Chasmata, in the neighborhood of Iseult and Tristram impact craters (Fig. 3). Despite the quality of the image and the crater coverage, which may bias the results, six generations of overlapping lineament group were identified, indicating changing orientation of the stress field, by the changing direction of the lineaments` strike. Such changing orientation may result from the asynchronous rotation, causing the change of the tidal dissipation pattern in time at one exact spot on the icy surface of Mimas. Furthermore, it also supports the continuously changing tidal influence and stress field orientation in the ice crust.

4. Conclusions

Despite the theories provided by simulations and models mentioned above, some geological evidence still left questions about the putative subsurface ocean hiding under the ice shell of Mimas. Its existence may be supported by the lineaments and their overlapping allocation with the tidal dissipation pattern, but there are some doubts about its juvenile state of development. In the agreement of a "newly" appeared subsurface ocean, the dominance of simple lineament types (undifferentiated linea, ridges, and troughs) suggests embryonic phase tectonic evolution. In contrast, the quasi-parallel, lineated, and ridged band and groove lane analog morphology, the features' relatively degraded and relaxed state, and their overlapping relation show a more mature state. Considering such observations, in the recent state of the geological/tectonic exploration of Mimas, it is difficult to decide that the appearance of the poorly developed, simple linear features is i) the result of their embryonic evolutionary phase, and they will develop further, as the putative young subsurface ocean evolves, or ii) their development paused in an early tectonic phase, and they fall in a dormant state or even stopped evolving, along with the development of the theoretical subsurface ocean.

Acknowledgments
*We want to thank the anonymous reviewers of this article for taking the time and effort to review the manuscript. We appreciate their valuable comments and suggestions, which helped improve this manuscript's quality.*

References

Bradák, B, Okumi, M. 2024. The Geological Map of Mimas v1.0-2023. Geosciences 14(1), 25. https://doi.org/10.3390/geosciences14010025

Bradák, B., Kimura, J., Kereszturi, Á., Gomez, C. 2023. Separating quasi-continuous and periodic components of lineament formation at the Belus – Phoenix - Rhadamanthys Linea "triangle" on Europa. Icarus 391, 115367, February 2023. https://doi.org/10.1016/j.icarus.2022.115367

Bruesch, L.S., Asphaug, E. 2004. Modeling global impact effects on middle-sized icy bodies: applications to Saturn's moons. Icarus 168(2), 457-466. https://doi.org/10.1016/j.icarus.2003.11.007

Croft, S.K., 1991. Mimas: Tectonic structure and geologic history. NASA, Washington, Reports of Planetary Geology and Geophysics Program, 1990. https://ntrs.nasa.gov/api/citations/19920001546/downloads/19920001546.pdf

Ćuk, M., Rhoden, A.R. 2024. Mimas's surprise ocean prompts an update of the rule book for moons. Nature 626, 263-264. https://doi.org/10.1038/d41586-024-00194-6

Dameron, A.C., Burr, D.M. 2018. Europan double ridge morphometry as a test of formation models. Icarus 305, 225–249. https://doi.org/10.1016/j.icarus.2017.12.009

Denton, C.A., Rhoden, A.R., 2022. Tracking the evolution of an ocean within Mimas using the Herschel impact basin. Geophysical Research Letters, 49(24), p.e2022GL100516. https://doi.org/10.1029/2022GL100516 - Herschel morphology and interior

Hay, H.C., Matsuyama, I. 2019. Nonlinear tidal dissipation in the subsurface oceans of Enceladus and other icy satellites. Icarus 319,68-85. https://doi.org/10.1016/j.icarus.2018.09.019

Howell, S.M., Pappalardo, R.T. 2018. Band formation and ocean-surface interaction on Europa and Ganymede. Geophys. Res. Lett. 45, 4701–4709. https://doi.org/10.1029/2018GL077594

Howell, S.M., Pappalardo, R.T. 2020. NASA's Europa Clipper—a mission to a potentially habitable ocean world. Nat Commun v. 11, 1311. https://doi.org/10.1038/s41467-020-15160-9

Lainey, V., Rambaux, N., Tobie, G. et al. 2024. A recently formed ocean inside Saturn's moon Mimas. Nature 626, 280–282 (2024) https://doi.org/10.1038/s41586-023-06975-9

Matsuyama, I., Beuthe, M., Hay, H.C., Nimmo, F., Kamata, S. 2018. Ocean tidal heating in icy satellites with solid shells. Icarus 312, 208-230. https://doi.org/10.1016/j.icarus.2018.04.013

Moore, J.M., Ahern, J.L. 1983. The geology of Tethys. Journal of Geophysical Research: Solid Earth, 88(S02), pp.A577-A584. https://doi.org/10.1029/JB088iS02p0A577

Moore, J.M., Schenk, P.M., Bruesch, L.S., Asphaug, E., McKinnon, W.B. 2004. Large impact features on middle-sized icy satellites. Icarus 171(2), 421-443. https://doi.org/10.1016/j.icarus.2004.05.009

Multhaup, K. and Spohn, T., 2007. Stagnant lid convection in the mid-sized icy satellites of Saturn. Icarus, 186(2), 420-435. https://doi.org/10.1016/j.icarus.2006.09.001

Neveu, M., Rhoden, A.R., 2017. The origin and evolution of a differentiated Mimas. Icarus, 296, 183-196. https://doi.org/10.1016/j.icarus.2017.06.011

Nimmo, F., Pappalardo, R. T. 2016. Ocean worlds in the outer solar system. J. Geophys. Res. Planets 121, 1378–1399. https://doi.org/10.1002/2016JE005081

Noyelles, B., 2017. Interpreting the librations of a synchronous satellite–How their phase assesses Mimas' global ocean. Icarus, 282, 276-289. https://doi.org/10.1016/j.icarus.2016.10.001

Prockter, L.M, Patterson, W. 2009. Morphology and evolution of Europa's ridges and bands. In Europa (eds. Pappalardo, R.T., McKinnon, W.B., Khurana, K.K.), University of Arizona Press, pp. 237-258 https://doi.org/10.2307/j.ctt1xp3wdw.16

Quillen, A.C., Giannella, D., Shaw, J.G., Ebinger, C. 2016. Crustal failure on icy moons from a strong tidal encounter. Icarus 275, 267–280. https://doi.org/10.1016/j.icarus.2016.04.003

Rhoden, A.R., Henning, W., Hurford, T.A., Patthoff, D.A., Tajeddine, R. 2017. The implications of tides on the Mimas ocean hypothesis. Journal of Geophysical Research: Planets, 122(2), 400-410. https://doi.org/10.1002/2016JE005097

Rhoden, A.R., Walker, M.E. 2022. The case for an ocean-bearing Mimas from tidal heating analysis. Icarus, 376, p.114872. https://doi.org/10.1016/j.icarus.2021.114872

Rhoden, A.R. 2023. Mimas: Frozen Fragment, Ring Relic, or Emerging Ocean World?. Annual Review of Earth and Planetary Sciences, 51, 367-387. https://www.annualreviews.org/doi/abs/10.1146/annurev-earth-031621-061221

Schenk, P.M. 1989. Crater formation and modification on the icy satellites of Uranus and Saturn: Depth/diameter and central peak occurrence. Journal of Geophysical Research: Solid Earth, 94(B4), 3813-3832. https://doi.org/10.1029/JB094iB04p03813

Schmedemann, N., Neukum, G. 2011. Impact crater size-frequency distribution (SFD) and surface ages on Mimas. In 42nd Annual Lunar and Planetary Science Conference (No. 1608, p. 2772). https://www.lpi.usra.edu/meetings/lpsc2011/pdf/2772.pdf

Stooke, P.J., 1989, March. Geology of Mimas. In Abstracts of the Lunar and Planetary Science Conference, volume 20, page 1069,(1989) (Vol. 20). https://adsabs.harvard.edu/full/1989LPI....20.1069S

Tajeddine, R., Rambaux, N., Lainey, V., Charnoz, S., Richard, A., Rivoldini, A., Noyelles, B., 2014. Constraints on Mimas' interior from Cassini ISS libration measurements. Science, 346(6207), 322-324. https://www.science.org/doi/full/10.1126/science.1255299